\documentclass[aps,prd,twocolumn,10pt,floatfix]{revtex4-2}

\usepackage{amsmath,amssymb,bm}
\usepackage{graphicx}
\usepackage{xcolor}
\usepackage{booktabs}
\usepackage{hyperref}

\begin{document}

\title{A halo-based intrinsic-alignment model for simulation-based inference}


\author{M.~Gatti}
\affiliation{Institut de Ci\`encies de l'Espai (ICE, CSIC),
Campus UAB, 08193 Barcelona, Spain}


\begin{abstract}

Simulation-based inference (SBI) and other field-level weak-lensing analyses require intrinsic-alignment (IA) models that generate realistic intrinsic-ellipticity fields across two-point and non-Gaussian observables. We develop a halo-based IA prescription with separate central and satellite components and test it against intrinsic galaxy shapes measured in the FLAMINGO hydrodynamical simulation. We jointly fit two-point and higher-order statistics involving the IA, lensing, and density fields, including correlations between the intrinsic shapes of distinct galaxies, and compare against field-level nonlinear-alignment (NLA) and density-weighted NLA prescriptions.

We first evaluate the models at the positions of the original FLAMINGO galaxies, fixing the galaxy--matter connection and isolating the IA response. NLA and $\delta$NLA leave substantial discrepancies on nonlinear scales and across higher-order observables, whereas the halo model provides a simultaneous description of the full set of statistics. We then remove the FLAMINGO galaxy catalogue and regenerate the source population from the halo field using a low-dimensional halo occupation model; the agreement is preserved even when the detailed occupation is only approximately reproduced. Much of this performance can be retained with a compact four-parameter portable model, and the same architecture transfers successfully to an independent gravity-only $N$-body realization with different halo definitions and shape measurements.

These results show that nonlinear field-level IA modelling can be substantially improved over simple NLA-like prescriptions without introducing a large nuisance-parameter space. The halo model therefore provides a promising route for incorporating intrinsic alignments into simulation-based weak-lensing inference, although further validation across halo masses, hydrodynamical simulations, and realistic survey selections will be needed before application to data.
\end{abstract}

\maketitle
\section{Introduction}
\label{sec:introduction}

Weak gravitational lensing provides a direct probe of the projected matter
distribution and has become one of the primary tools for constraining the
growth of cosmic structure and the late-time expansion of the Universe
\citep{Bartelmann2001,Kilbinger2015,Mandelbaum2018}. The new
generation of wide and deep imaging surveys, including Euclid
\citep{Laureijs2011,Mellier2025}, the Vera C.\ Rubin Observatory Legacy
Survey of Space and Time (LSST; \citep{LSST2009,Ivezic2019}), and
the Nancy Grace Roman Space Telescope \citep{Akeson2019}, will dramatically increase the
statistical precision of cosmic-shear measurements. Exploiting this
precision requires a correspondingly accurate description of astrophysical
and observational effects that modify the measured galaxy-shape field.

One of the most important astrophysical contributions is the intrinsic
alignment (IA) of galaxies: correlations between galaxy shapes and the
surrounding large-scale structure that arise from galaxy formation and
evolution rather than gravitational lensing
\citep{Heavens2000,Catelan2001,Hirata2004}. IA contributes both
intrinsic--intrinsic (II) and gravitational--intrinsic (GI) correlations to
cosmic-shear measurements and, if incorrectly modelled, can bias cosmological
constraints. At the same time, IA is not merely a nuisance signal. Because
galaxy orientations respond to the tidal field and to the assembly of their
host haloes, IA contains information about galaxy formation, the
galaxy--halo connection, and the large-scale structure itself
\citep{Joachimi2015,Kiessling2015,Kirk2015,Lamman2024,Chisari2025}.
Accurately describing IA is therefore important both for mitigating its
impact on weak-lensing cosmology and for exploiting its astrophysical and
cosmological information.

Several approaches have been developed to model this signal. On sufficiently large scales, tidal-alignment models relate galaxy shapes to the surrounding tidal field. The nonlinear alignment (NLA) model \citep{Hirata2004,Bridle2007} provides a widely used phenomenological extension into the nonlinear regime, while higher-order tidal responses can be included through models such as tidal alignment and tidal torquing (TATT) \citep{Blazek2019}. An alternative description is provided by halo models, in which the IA field is built from the shapes, orientations, occupation, and spatial distribution of central and satellite galaxies within dark-matter haloes \citep{Schneider2010,Fortuna2021}. Closely related empirical HOD+IA frameworks have been developed to generate nonlinear alignment mocks and reproduce position--orientation and orientation--orientation correlations in hydrodynamical simulations \citep{VanAlfen2024}, with recent extensions using neural-network emulation and differentiable HOD+IA modelling for efficient inference \citep{Pandya2025IAEmu,Pandya2026diffHODIA}. Related prescriptions have also been used to populate large $N$-body lightcones with intrinsic galaxy shapes \citep{Hoffmann2022,EuclidFlagshipIA2026}. These approaches need not remain equivalent on nonlinear scales, where the galaxy--halo connection and intra-halo structure become increasingly important.

This issue becomes particularly relevant as weak-lensing analyses move beyond two-point statistics. Higher-order moments, peaks and minima, wavelet-based summaries, topological statistics, and learned map-level statistics can recover information from the nonlinear matter field that is absent from the shear power spectrum alone. These observables are increasingly analysed with simulation-based methods \citep{ZorrillaMatilla2020,Sabyr2022,Ajani2023,Vinciguerra2026,
HarnoisDeraps2022,Gatti2024a,Gatti2025,Jeffrey2025,Zeghal2025}, which require IA to be propagated at field level. Agreement with a two-point IA amplitude is then not sufficient: different prescriptions can produce similar power spectra while generating different non-Gaussian structure \citep{HarnoisDeraps2026}.

Hydrodynamical simulations provide a natural test bed because galaxy shapes,
halo properties, and the matter distribution are generated self-consistently. They have established strong dependencies of IA on galaxy type, stellar and halo mass, redshift, shape definition, and galaxy--halo misalignment \citep{Velliscig2015,Chisari2017,Joachimi2015,Kiessling2015}. Recent large-volume simulations have extended these studies to nonlinear and map-level statistics: MillenniumTNG has been used to study galaxy--halo alignments and full lightcone IA maps \citep{Delgado2023,Ferlito2025}, IllustrisTNG to quantify IA effects on non-Gaussian lensing statistics \citep{Lee2026}, and FLAMINGO to measure the mass dependence of two- and three-point alignment statistics \citep{Herle2026,Vedder2026}.

In this work, we develop and test a halo-based IA prescription for field-level weak-lensing forward modelling. Building on earlier halo-based approaches \citep{Schneider2010,Fortuna2021,Hoffmann2022,VanAlfen2024,Pandya2025IAEmu,Pandya2026diffHODIA,HarnoisDeraps2026}, we ask whether a low-dimensional halo model can reproduce the IA field of a hydrodynamical simulation simultaneously across two-point and non-Gaussian statistics. Using FLAMINGO as the target, we construct the IA field from halo properties, occupations, and central/satellite geometry, and compare it with NLA and density-weighted NLA across several galaxy samples and summary statistics.

A second goal is to test whether the same construction can be transferred to gravity-only simulations, as required for simulation-based inference (SBI). We therefore assess both the portability of the source-population model and the robustness of the IA prescription across independent simulation realizations.

The paper is organized as follows. In
Sec.~\ref{sec:simulation_inputs} we describe the simulation inputs and
galaxy samples. Section~\ref{sec:ia_prescriptions} introduces the NLA,
$\delta$NLA, and halo-based IA prescriptions.
In Sec.~\ref{sec:summary_statistics} we define the summary statistics and
model-fitting procedure. The main results are presented in
Sec.~\ref{sec:results}, followed by their interpretation and implications
for field-level forward modelling in Sec.~\ref{sec:discussion}.
We summarize our conclusions in Sec.~\ref{sec:summary}. Additional results
for the colour-selected samples are given in
Appendix~\ref{app:sample_results}.

\section{Simulation inputs and galaxy populations}
\label{sec:simulation_inputs}

\subsection{FLAMINGO simulation and lightcone}
\label{sec:flamingo_sim}

Our primary analysis uses the fiducial hydrodynamical \texttt{L1\_m8} realization of the FLAMINGO simulation suite \citep{Schaye2023}. The simulation evolves a periodic comoving volume of side length $1\,{\rm Gpc}$ with $3600^3$ gas particles, the same number of cold-dark-matter particles, and $2000^3$ massive-neutrino particles. The initial gas and cold-dark-matter particle masses are $1.34\times10^8\,M_\odot$ and $7.06\times10^8\,M_\odot$, respectively. FLAMINGO was evolved with \textsc{swift} \citep{Schaller2024} and includes radiative cooling, star formation, stellar evolution and feedback, and black-hole growth and feedback. The feedback parameters were calibrated against the low-redshift galaxy stellar mass function and gas fractions of groups and clusters \citep{Kugel2023}; intrinsic-alignment observables were not used in this calibration.

The simulation adopts the fiducial FLAMINGO D3A flat-$\Lambda$CDM cosmology, with $h=0.681$, $\Omega_m=0.306$, $\Omega_b=0.0486$, $\Omega_\Lambda=0.694$, $\sigma_8=0.807$, $n_s=0.967$, and $\sum m_\nu=0.06\,{\rm eV}$. We use the full-sky FLAMINGO lightcone over $0\leq z<2$, where the resolved galaxy population remains sufficiently dense for the shape measurements used below. Galaxy positions and redshifts are taken from the lightcone catalogue, while galaxy and halo properties are obtained from the associated SOAP catalogues \citep{McGibbon2025}.

\subsection{Stellar tensors and projected intrinsic ellipticities}
\label{sec:stellar_projection}

We measure intrinsic galaxy shapes from the stellar mass distribution of resolved FLAMINGO galaxies. We require at least 100 bound stellar particles per galaxy and retain only objects with finite lightcone coordinates and well-defined stellar shape tensors. We use the iterative, unreduced \texttt{BoundSubhalo/StellarInertiaTensor} quantity provided by SOAP \citep{McGibbon2025}, corresponding to the mass-weighted second moment of the stellar distribution,
\begin{equation}
S_{ij} =
\frac{\sum_{k\in\mathcal{A}} m_k x_{k,i} x_{k,j}}
{\sum_{k\in\mathcal{A}} m_k},
\label{eq:stellar_tensor}
\end{equation}
where $\boldsymbol{x}_k$ is the position of stellar particle $k$ relative to the galaxy centre, $m_k$ is its mass, and $\mathcal{A}$ denotes the set of particles entering the iterative shape measurement.

We project each three-dimensional stellar tensor onto the local plane of the sky. For a galaxy at right ascension $\alpha$ and declination $\delta$, the local east and north unit vectors are
\begin{equation}
\hat{\boldsymbol e}_{E}=(-\sin\alpha,\cos\alpha,0),
\end{equation}
and
\begin{equation}
\hat{\boldsymbol e}_{N}=(-\sin\delta\cos\alpha,-\sin\delta\sin\alpha,\cos\delta).
\end{equation}
The projected tensor is then
\begin{equation}
Q_{ab}=\hat{\boldsymbol e}_{a}^{\mathsf T}S\hat{\boldsymbol e}_{b},\qquad a,b\in\{E,N\},
\label{eq:stellar_tensor_projection}
\end{equation}
and we convert it to the complex intrinsic ellipticity
\begin{equation}
\epsilon \equiv e_1 + \mathrm{i}e_2 =
\frac{Q_{EE}-Q_{NN}+2\mathrm{i}Q_{EN}}
{Q_{EE}+Q_{NN}+2\sqrt{Q_{EE}Q_{NN}-Q_{EN}^2}}.
\label{eq:stellar_epsilon}
\end{equation}

These shapes trace the projected stellar mass distribution of the simulated galaxies and are not intended to reproduce a specific observational shape estimator: they are not luminosity weighted and do not include PSF convolution, image noise, blending, or observational shape-selection effects.

\subsection{Halo quantities}
\label{sec:halo_inputs}

The halo-based IA model uses the FLAMINGO SOAP halo catalogues independently of the selected galaxy catalogue. We retain haloes with $M_{200c}\geq10^{12}\,M_\odot$, finite lightcone positions, and valid dark-matter shape tensors. Here $M_{200c}$ is taken from \texttt{SO/200\_crit/TotalMass} and denotes the total mass enclosed within a sphere of mean density $200\rho_{\rm crit}(z)$.

The dark-matter shape tensors provide the orientation information used by the halo model. We project them onto the local plane of the sky using the same procedure as for the stellar tensors in Sec.~\ref{sec:stellar_projection}. The stellar tensors define the target galaxy shapes, whereas the dark-matter tensors provide the structural information available to the model.

Galaxy and halo selections are kept separate throughout the analysis. The model can therefore use all eligible haloes rather than only the known hosts of galaxies in the target sample. Host identifiers are used only in the matched tests and to characterize the galaxy--halo connection.

\subsection{Galaxy samples and redshift distributions}
\label{sec:galaxy_samples}

\begin{figure}
    \centering
    \includegraphics[width=0.45 \textwidth]{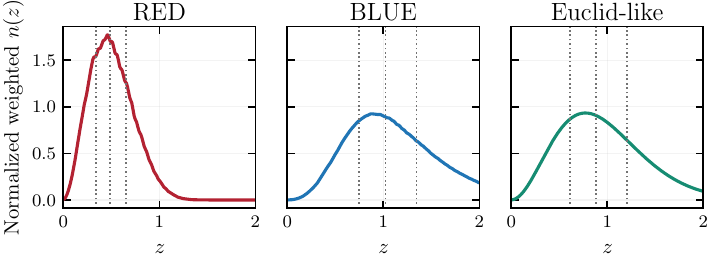}
    \caption{
    Weighted redshift distributions of the RED, BLUE, and Euclid-like
    galaxy samples. Vertical dotted lines mark the boundaries of the four
    equal-weight tracer bins used to construct the projected density and
    lensing fields. The RED and BLUE samples retain their native FLAMINGO
    distributions, while the Euclid-like sample is obtained by reweighting
    the full resolved population to Eq.~\eqref{eq:euclid_nz}.
    }
    \label{fig:sample_nz}
\end{figure}

\begin{table}[t]
\centering
\begin{tabular}{lrrr}
\toprule
Sample & $N_{\rm gal}$ & $\langle z\rangle_w$ & $f_{\rm sat,w}$\\
\midrule
RED         & 2,503,589  & 0.506 & 0.197\\
BLUE        & 10,691,593 & 1.055 & 0.081\\
Euclid-like & 13,195,182 & 0.925 & 0.105\\
\bottomrule
\end{tabular}
\caption{
Resolved FLAMINGO galaxy samples over $0\leq z<2$. The RED and BLUE
samples retain their native redshift distributions, whereas the Euclid-like
sample contains all colours and is reweighted according to
Eq.~\eqref{eq:euclid_nz}. Mean redshifts and satellite fractions include
the corresponding object weights.
}
\label{tab:flamingo_current_samples}
\end{table}

We consider three galaxy samples. The first two split the resolved FLAMINGO population by rest-frame colour,
\begin{equation}
\mathrm{RED}:~g-r\geq0.6,\qquad
\mathrm{BLUE}:~g-r<0.6,
\end{equation}
using the dust-free rest-frame GAMA-band luminosities provided by SOAP, with $g-r=-2.5\log_{10}(L_g/L_r)$. Both samples contain centrals and satellites and retain their native FLAMINGO redshift distributions. The colour split is used to select populations with different IA behaviour; we impose no additional morphology or star-formation-rate cuts.

We also define an all-colour sample with a redshift distribution chosen to resemble that of a Stage-IV weak-lensing source catalogue,
\begin{equation}
n_{\rm E}(z)\propto z^2\exp\left[-(z/z_0)^{3/2}\right],\qquad z_0=0.636,
\label{eq:euclid_nz}
\end{equation}
following the commonly adopted Euclid-like source distribution \citep{Laureijs2011,EuclidForecast2020}. We refer to this population as the \emph{Euclid-like} sample. The weights are given by the ratio of the target and parent redshift distributions and normalized to unit mean over the populated redshift range.

Figure~\ref{fig:sample_nz} shows the weighted redshift distributions of the three samples together with the boundaries of the four tracer bins used below. The RED and BLUE samples retain their native FLAMINGO distributions, whereas the Euclid-like sample is shifted towards the target source distribution.

We stress that the Euclid-like sample is defined only through its redshift weighting. We do not impose Euclid VIS magnitude, size, completeness, or shape-measurement selections, and the requirement of a resolved stellar tensor preferentially selects better-resolved and typically more massive galaxies and host haloes. None of the three samples should therefore be interpreted as a detailed mock of an observed source population.

Table~\ref{tab:flamingo_current_samples} summarizes the basic properties of the three samples.

\subsection{IA maps and external tracers}
\label{sec:maps_tracers}

We pixelize the intrinsic galaxy shapes into full-sky HEALPix maps with $N_{\rm side}=2048$. For each sample, the IA field in pixel $p$ is
\begin{equation}
I_p=\frac{\sum_{i\in p} w_i\,\epsilon_i}{\sum_{i\in p} w_i},
\end{equation}
where $w_i=1$ for the RED and BLUE samples and is given by the fixed redshift weight for the Euclid-like sample. At this resolution the IA maps are sparse: the RED, BLUE, and Euclid-like catalogues contain numbers of galaxies corresponding to about $5.0\%$, $21.4\%$, and $26.4\%$ of the total number of pixels, respectively, and therefore retain substantial finite-sampling and intrinsic-shape noise.

We complement these sparse IA maps with dense full-sky matter and lensing fields from the FLAMINGO lightcone. For each galaxy sample, we divide the weighted source distribution into four tracer bins containing equal total analysis weight. The bin edges are therefore sample dependent and correspond to weighted redshift quartiles. Within tracer bin $t$, the projected matter field is
\begin{equation}
d_t(\hat{\boldsymbol n})=\sum_s w_{ts}\,\delta_s(\hat{\boldsymbol n}),
\end{equation}
where $\delta_s$ is the matter overdensity in radial shell $s$ and $w_{ts}$ is the corresponding source-distribution weight. The lensing fields are constructed from the same density shells with the GLASS projection framework \citep{Tessore2023}; we denote their scalar E-mode by $G_t$.

No observational noise is added to $d_t$ or $G_t$. Although both are derived from the same matter field, they probe it with different radial kernels: $d_t$ follows the source distribution in each tracer bin, whereas $G_t$ integrates foreground structure with the lensing kernel. Statistics containing a single IA factor, including $dI$, $GI$, and mixed higher-order moments, therefore contain shape noise only through the IA field and provide high-signal-to-noise probes of its redshift and environmental dependence without further subdividing the sparse source sample.

\subsection{Independent PKDGRAV portability realization}
\label{sec:pkdgrav_sim}

To test whether the halo-based construction can be transferred beyond the hydrodynamical realization used for calibration, we use an independent dark-matter-only simulation evolved with \textsc{PKDGRAV3} \citep{Potter2017}. \textsc{PKDGRAV3} is a widely used $N$-body code that has been employed, for example, in the production of the Euclid Flagship simulation \citep{EuclidFlagship2025}, as well as in the Gower Street simulation suites used for recent weak-lensing simulation-based inference analyses of DES data \citep{Gatti2024a,Gatti2025,Jeffrey2025}. This makes it particularly well suited for testing whether the IA prescription can be transferred to the type of gravity-only simulations used in survey-level forward-modelling analyses. The realization used here follows the same numerical and lightcone pipeline, but is run at the fiducial FLAMINGO D3A cosmology. The periodic box has side length $1250\,h^{-1}{\rm Mpc}$ and contains $1350^3$ dark-matter particles. We construct the full-sky matter and halo lightcone quantities required by the IA model and apply the same source-population and IA prescriptions without using any FLAMINGO galaxy positions or host identities.

The PKDGRAV halo catalogue differs from the FLAMINGO SOAP catalogue in both definition and measured properties. PKDGRAV uses friends-of-friends (FoF) haloes with FoF masses and inertia tensors, whereas FLAMINGO uses SOAP spherical-overdensity haloes selected by $M_{200c}$. The transfer test therefore changes not only the density realization, but also the halo population, mass definition, and shape measurements available to the model.

The projected density and lensing fields are constructed using the same radial binning and projection procedure adopted for FLAMINGO. Since PKDGRAV is dark-matter-only, we approximately account for baryonic effects by rescaling the density modes in each radial shell using the scale-dependent ratio of the FLAMINGO hydrodynamical and dark-matter-only matter power spectra. The lensing fields are then generated from these corrected shells with the same GLASS projection used above.

\section{Field-level IA prescriptions}
\label{sec:ia_prescriptions}

We compare three prescriptions for the coherent intrinsic-alignment (IA)
field: a field-level nonlinear-alignment model (NLA), its density-weighted
extension ($\delta$NLA), and a halo-based model with separate central and
satellite contributions. In all cases, we distinguish between the IA
response itself and the galaxy population used to sample it. The latter
encodes the galaxy--matter connection, determining where galaxies are
located relative to the structures that generate the IA signal and,
consequently, which parts of the underlying response enter the measured
shape map. This distinction is particularly important for the sparse galaxy samples considered here, since an inaccurate source-sampling prescription can misweight the regions that dominate the IA signal. We evaluate all three models separately for the RED, BLUE, and Euclid-like populations.

We consider two complementary implementations of this sampling. In the
\emph{matched} case, all three IA prescriptions are evaluated at the
positions of the original FLAMINGO galaxies. No source population is
generated, and the comparison primarily tests the IA response conditional
on the galaxy--matter connection realized in the hydrodynamical simulation.
In the \emph{portable} case, the FLAMINGO galaxy positions are not used and
the source population must also be modelled.

\subsection{Tidal-field models: NLA and
\texorpdfstring{$\delta$NLA}{delta-NLA}}
\label{sec:tidal_ia}

The NLA model relates intrinsic galaxy shapes to the large-scale tidal field, with the nonlinear matter density replacing the linear density field in the standard alignment prescription \citep{Bridle2007}. We implement it directly at field level using the projected FLAMINGO matter shells. For shell $r$, let $\delta_r(\hat{\boldsymbol n})$ denote the projected matter-density contrast and ${\cal T}_r$ the corresponding spin-2 tidal field. If $\delta_{r,\ell m}$ are the spherical-harmonic coefficients of the projected density field, we define
\begin{equation}
E^{{\cal T}}_{\ell m}
=
-\sqrt{\frac{(\ell-1)(\ell+2)}{\ell(\ell+1)}}
\frac{P^{(2)}_\ell}{P^{(0)}_\ell}
\delta_{r,\ell m},
\qquad
B^{{\cal T}}_{\ell m}=0,
\label{eq:tidal_shell_operator}
\end{equation}
for $\ell\geq2$, with the monopole and dipole set to zero. Here $P^{(0)}_\ell$ and $P^{(2)}_\ell$ are the scalar and spin-2 HEALPix pixel window functions. This construction should be understood as a projected-shell realization of the tidal-alignment field rather than an explicit reconstruction of the three-dimensional tidal tensor at each galaxy position.

For a galaxy $i$ in shell $r(i)$, we define
\begin{equation}
U_i=-F_1(z_i)\,{\cal T}_{r(i)}(\hat{\boldsymbol n}_i),
\label{eq:nla_unit_response}
\end{equation}
with
\begin{equation}
F_1(z)=-\frac{C_1\rho_{\rm crit,0}\Omega_m}{D(z)},
\qquad
C_1=5\times10^{-14}\,h^{-2}M_\odot^{-1}\mathrm{Mpc}^{3},
\end{equation}
where $D(z)$ is the linear growth factor normalized to $D(0)=1$. The NLA response is then
\begin{equation}
m_i^{\rm NLA}
=
A_0
\left(\frac{1+z_i}{1+z_p}\right)^\eta
U_i,
\label{eq:nla_object_response}
\end{equation}
with pivot redshift $z_p=0.67$ and free parameters $A_0$ and $\eta$.

We also consider a density-weighted extension,
\begin{align}
m_i^{\delta{\rm NLA}}
={}&
\Bigg[
A_0
\left(\frac{1+z_i}{1+z_p}\right)^\eta
+
A_{\delta,0}
\left(\frac{1+z_i}{1+z_p}\right)^{\eta_\delta}
\delta_{r(i)}(\hat{\boldsymbol n}_i)
\Bigg]U_i,
\label{eq:dnla_object_response}
\end{align}
which adds a density-weighted amplitude $A_{\delta,0}$ and redshift dependence $\eta_\delta$.

This $\delta$NLA construction is motivated by density-weighted tidal-alignment terms appearing in perturbative IA expansions \citep{Blazek2019}, but it is not intended to reproduce perturbative TATT exactly. Both the density and tidal fields are built from the fully nonlinear matter field, and their product is formed only after projection into finite radial shells rather than from the corresponding three-dimensional composite operator. We therefore treat $\delta$NLA as a phenomenological field-level extension of NLA; in particular, it contains no independent quadratic tidal-torquing contribution.

In the matched construction, both NLA and $\delta$NLA are evaluated directly at the positions and redshifts of the selected FLAMINGO galaxies, preserving the galaxy--matter connection of the hydrodynamical simulation and isolating differences in the IA response. We also tested a fully synthetic implementation in which source positions are generated from the projected density field using density-dependent Poisson sampling, but for the sparse shape-selected samples considered here this prescription does not reproduce the galaxy--matter connection with sufficient accuracy. We therefore use the matched construction for the tidal-field comparisons in the main analysis.

\begin{table*}
\centering
\caption{Free parameters of the IA prescriptions used in this work. The matched halo model uses the FLAMINGO galaxy population directly and therefore contains only the six IA-response parameters, while the portable model additionally varies the two source-population parameters $f_c$ and $f_{\rm sat}$. Quantities such as $M_{\min}(z)$, $A_{{\rm pop},s}(z)$, $q_h$, and $\phi_h$ are derived from the abundance constraints or halo catalogue and are not free parameters.}
\label{tab:ia_model_parameters}
\begin{tabular}{llll}
\toprule
Model & Parameter & Meaning & Exploration interval \\
\midrule
NLA
& $A_0$ & IA amplitude at $z_p=0.67$ & $(-\infty,\infty)$ \\
& $\eta$ & IA redshift evolution & $[-6,6]$ \\
\midrule
$\delta$NLA
& $A_0$ & Tidal-alignment amplitude at $z_p=0.67$ & $(-\infty,\infty)$ \\
& $\eta$ & Tidal-alignment redshift evolution & $[-6,6]$ \\
& $A_{\delta,0}$ & Density-weighted IA amplitude at $z_p=0.67$ & $(-\infty,\infty)$ \\
& $\eta_\delta$ & Density-weighted redshift evolution & $[-6,6]$ \\
\midrule
Halo, matched
& $A_{c,0}$ & Central IA amplitude at $z_p=0.67$ & $[0,2]$ \\
& $\eta_c$ & Central redshift evolution & $[-6,6]$ \\
& $b_{{\rm TA},c}$ & Central density response & $[-3,3]$ \\
& $A_{s,0}$ & Satellite IA amplitude at $z_p=0.67$ & $[0,0.5]$ \\
& $\eta_s$ & Satellite redshift evolution & $[-6,6]$ \\
& $b_{{\rm TA},s}$ & Satellite density response & $[-3,3]$ \\
\midrule
Halo, portable
& $f_c$ & High-mass central completeness & $[0.05,1]$ \\
& $f_{\rm sat}$ & Satellite fraction in each redshift cell & $[0,0.4]$ \\
& $A_{c,0}$ & Central IA amplitude at $z_p=0.67$ & $[0,2]$ \\
& $\eta_c$ & Central redshift evolution & $[-6,6]$ \\
& $b_{{\rm TA},c}$ & Central density response & $[-3,3]$ \\
& $A_{s,0}$ & Satellite IA amplitude at $z_p=0.67$ & $[0,0.5]$ \\
& $\eta_s$ & Satellite redshift evolution & $[-6,6]$ \\
& $b_{{\rm TA},s}$ & Satellite density response & $[-3,3]$ \\
\bottomrule
\end{tabular}
\end{table*}

\subsection{Halo-based IA model}
\label{sec:halo_ia}

Our third prescription associates the IA signal directly with dark-matter haloes, following the general picture of halo-based IA models \citep{Schneider2010,Fortuna2021}. We treat central and satellite galaxies separately and model three ingredients: the source population occupying each halo, the projected halo geometry that sets the central template and satellite spatial distribution, and separate IA responses for centrals and satellites.

We consider two implementations. In the \emph{matched} construction, we use the FLAMINGO galaxies directly, including their central/satellite classification, host assignments, and measured positions. In the \emph{portable} construction, these quantities are unavailable and the source population is generated from the halo catalogue.

\paragraph{Portable source population.}

The portable model is conditioned on the target redshift abundance while retaining a minimal halo-mass dependence. In each redshift cell, the mean central and satellite occupations are
\begin{align}
\overline N_c(M,z)
&=
\frac{f_c}{2}
\left[
1+\operatorname{erf}
\left(
\frac{\log_{10}M-\log_{10}M_{\min}(z)}
{\sigma_{\log M}}
\right)
\right],\\
\overline N_s(M,z)
&=
A_{{\rm pop},s}(z)\,
\overline N_c(M,z)
\left(
\frac{M}{M_{\min}(z)}
\right)^\alpha ,
\label{eq:portable_minimal_hod}
\end{align}
with $\sigma_{\log M}=0.2$ and $\alpha=1$ fixed in the fiducial model.

For a target source count $N^{\rm target}_j$ in redshift cell $j$ and satellite fraction $f_{\rm sat}$, we impose
\begin{equation}
N^{\rm target}_{c,j}=(1-f_{\rm sat})N^{\rm target}_j,
\qquad
N^{\rm target}_{s,j}=f_{\rm sat}N^{\rm target}_j,
\label{eq:portable_target_split}
\end{equation}
and determine $M_{\min}(z_j)$ and $A_{{\rm pop},s}(z_j)$ from
\begin{equation}
\sum_{h\in j}\overline N_{c,h}=N^{\rm target}_{c,j},
\qquad
\sum_{h\in j}\overline N_{s,h}=N^{\rm target}_{s,j}.
\label{eq:portable_hod_constraints}
\end{equation}
These quantities are therefore derived rather than free parameters, leaving only $f_c$ and $f_{\rm sat}$ to control the source population. This deliberately low-dimensional construction is intended to capture the broad halo-mass and redshift dependence of the selected population without introducing a detailed halo occupation distribution (HOD) parameterization.
Centrals and satellites are then drawn as
\begin{equation}
N_{c,h}\sim\operatorname{Bernoulli}(\overline N_{c,h}),
\qquad
N_{s,h}\sim\operatorname{Poisson}(\overline N_{s,h}).
\end{equation}
The satellite draw is independent of the realized central draw, so selected satellites can occur in haloes without a selected central. Centrals are placed at the halo centre, while satellite positions are generated from the spatial model described below.

\paragraph{Projected halo geometry.}

For each halo, we project the three-dimensional dark-matter shape tensor onto the local plane of the sky using the same procedure as for the stellar tensor in Sec.~\ref{sec:stellar_projection}. From the resulting two-dimensional tensor we define the complex projected halo ellipticity $\epsilon_h=e_{1,h}+{\rm i}e_{2,h}$ using Eq.~\eqref{eq:stellar_epsilon}.

For a central galaxy, we take the IA template to be the projected halo ellipticity,
\begin{equation}
u_{c,h}=\epsilon_h .
\label{eq:central_ia_geometry}
\end{equation}
The halo tensor therefore sets the preferred orientation and relative halo-to-halo variation of the central template, while the overall coherent normalization is absorbed by the central IA amplitude introduced below.

For satellites, we model both their spatial distribution within the host halo and their intrinsic orientation. Satellite positions are drawn from an elliptical projected-NFW profile whose orientation and axis ratio are inherited directly from the projected dark-matter halo tensor. Let $q_h$ and $\phi_h$ denote the projected halo axis ratio and major-axis position angle inferred from the tensor eigenvalues and eigenvectors. In coordinates aligned with the projected major axis, we draw a dimensionless projected radius $u$ from the projected NFW distribution and an azimuthal angle $\varphi$ uniformly in $[0,2\pi)$, with
\begin{equation}
X = R_{200c}^{\rm com}\,u\cos\varphi,
\qquad
Y = q_h R_{200c}^{\rm com}\,u\sin\varphi .
\end{equation}
Thus the anisotropy of the satellite distribution is fixed by the host-halo geometry and introduces no additional free parameter.

The comoving halo radius is
\begin{equation}
R_{200c}^{\rm com}
=
(1+z)
\left[
\frac{3M_{200c}}
{4\pi\,200\rho_{\rm crit}(z)}
\right]^{1/3},
\end{equation}
and the concentration follows the median $c_{200c}(M,z)$ relation of \citet{DiemerJoyce2019}, evaluated with \textsc{Colossus} \citep{Diemer2018Colossus}. We truncate the projected profile at $R_{200c}$ and sample
\begin{equation}
p(u)\propto u\,f_{\rm pNFW}(c_{200c}u),
\end{equation}
where $f_{\rm pNFW}$ is the projected NFW surface-density profile.

The halo-aligned offsets are rotated into the local east--north basis,
\begin{align}
d_E &= X\cos\phi_h-Y\sin\phi_h,\\
d_N &= X\sin\phi_h+Y\cos\phi_h .
\end{align}
We then assign each satellite a radial intrinsic alignment with respect to the host centre,
\begin{equation}
u_s=
\frac{(d_E+{\rm i}d_N)^2}
{d_E^2+d_N^2},
\label{eq:satellite_ia_geometry}
\end{equation}
where $|u_s|=1$ and the coherent satellite ellipticity amplitude is supplied separately by the IA response.

\paragraph{Central and satellite IA responses.}

For each galaxy, the coherent IA contribution is
\begin{align}
{\rm IA}_i^{\rm halo}
={}&
A_{c,0}
\left(\frac{1+z_i}{1+z_p}\right)^{\eta_c}
\left[1+b_{{\rm TA},c}\delta_i(z_i)\right]u_{c,i}
\nonumber\\
&+
A_{s,0}
\left(\frac{1+z_i}{1+z_p}\right)^{\eta_s}
\left[1+b_{{\rm TA},s}\delta_i(z_i)\right]u_{s,i},
\label{eq:halo_object_response}
\end{align}
where $u_c=0$ for satellites and $u_s=0$ for centrals. The six IA-response parameters are therefore
\begin{equation}
\left\{
A_{c,0},\eta_c,b_{{\rm TA},c},
A_{s,0},\eta_s,b_{{\rm TA},s}
\right\}.
\end{equation}
The amplitudes $A_{c,0}$ and $A_{s,0}$ set the coherent central and satellite responses at the pivot redshift $z_p=0.67$, while $\eta_c$ and $\eta_s$ describe their redshift evolution and $b_{{\rm TA},c}$ and $b_{{\rm TA},s}$ their dependence on the local density field.

In the matched construction, the FLAMINGO galaxy positions, central--satellite classifications, and host assignments are used directly, so only the six IA-response parameters are varied. The central or satellite response is selected according to the galaxy classification: centrals use the projected ellipticity of their host halo, while satellites use the radial template defined by their measured position relative to the host centre.

In the portable construction, the source catalogue and satellite positions are generated from the halo catalogue using the prescriptions above. The same six IA-response parameters are then applied to the synthetic sources, together with the two population parameters $f_c$ and $f_{\rm sat}$, for a total of eight free parameters. The quantities $M_{\min}(z)$ and $A_{{\rm pop},s}(z)$ are fixed by the abundance constraints, while $q_h$ and $\phi_h$ are measured directly from the projected halo tensor. The free parameters and exploration intervals of all IA prescriptions are summarized in Table~\ref{tab:ia_model_parameters}.

The use of projected halo shape as the central template is motivated by simulations and observations showing a statistical alignment between central galaxies and their host haloes, with substantial scatter \citep[e.g.][]{Wang2008,Shao2016}. We therefore interpret the halo tensor as a coherent template rather than a direct prediction of the stellar ellipticity: its orientation and relative halo-to-halo variation are retained, while the overall response is absorbed by $A_{c,0}$. For satellites, we adopt the standard picture of preferential radial alignment towards the host centre \citep{Schneider2010,Fortuna2021}.

\section{Summary statistics and IA calibration methodology}
\label{sec:summary_statistics}

We test the IA prescriptions using a common set of two-point and non-Gaussian summary statistics, which are fitted jointly in the calibration. The statistics include observables containing either one or two factors of the intrinsic-alignment field, together with the external density and lensing fields. The same summary operators are applied to FLAMINGO and to all forward-model realizations. Statistics containing two IA factors require a distinct-object estimator to remove same-galaxy self-pairs and are therefore treated separately at the measurement level, as described below.

All maps are constructed at \textsc{HEALPix} $N_{\rm side}=2048$ and band-limited at $\ell_{\max}=4096$. Let $I$ denote the scalar E-mode of the spin-2 intrinsic-ellipticity field, and let $G_t$ and $d_t$ denote the lensing E-mode and projected matter-density field in tracer bin $t=1,\ldots,4$. The IA field is constructed from the full selected source sample rather than split tomographically; the tracer index therefore labels only the lensing or density kernel. Before computing higher-order statistics, we subtract the mean of each field.

\paragraph{Two-point statistics.}

The main two-point summaries are
\begin{equation}
C_\ell^{G_t I}=\left\langle G_{t,\ell m} I^*_{\ell m}\right\rangle_m,
\qquad
C_\ell^{d_t I}=\left\langle d_{t,\ell m} I^*_{\ell m}\right\rangle_m,
\label{eq:GI_dI_spectra}
\end{equation}
which we refer to as $GI$ and $dI$, respectively. We bin the spectra using mode-count weighting,
\begin{equation}
\widehat C_b^{XI}=
\frac{\sum_{\ell\in b}(2\ell+1)\widehat C_\ell^{XI}}
{\sum_{\ell\in b}(2\ell+1)},
\qquad X\in\{G_t,d_t\}.
\label{eq:bandpower_estimator}
\end{equation}
The full measurement contains 29 bands over $2\leq\ell\leq4096$. The fiducial calibration uses the 14 bands with effective centres in $200\leq\ell\leq4096$, while lower multipoles are retained only as a large-scale consistency check.

\paragraph{One-IA higher-order statistics.}

For each tracer bin we measure five zero-lag moments containing one factor of the IA field,
\begin{align}
M_{G^2I,t}&=\left\langle G_t^2 I\right\rangle,
&M_{d^2I,t}&=\left\langle d_t^2 I\right\rangle,\nonumber\\
M_{GdI,t}&=\left\langle G_t d_t I\right\rangle,
&M_{G^3I,t}&=\left\langle G_t^3 I\right\rangle,
&M_{d^3I,t}&=\left\langle d_t^3 I\right\rangle.
\label{eq:one_I_moments}
\end{align}
These statistics probe nonlinear couplings between IA, lensing, and density that are not captured by the two-point functions \citep{Gatti2024a,Jeffrey2025,Gatti2024b}.

We also measure the mean IA signal at extrema selected in the external fields. For $X_t\in\{G_t,d_t\}$, maxima and minima above threshold $\nu$ are defined as
\begin{align}
{\cal P}^{+}_{X_t}(\nu)
&=
\left\{
p:\ p\ {\rm is\ a\ local\ maximum},\quad
X_t(p)\geq\nu\sigma_{X_t}
\right\},\nonumber\\
{\cal P}^{-}_{X_t}(\nu)
&=
\left\{
p:\ p\ {\rm is\ a\ local\ minimum},\quad
X_t(p)\leq-\nu\sigma_{X_t}
\right\},
\label{eq:extrema_definition}
\end{align}
where $\sigma_{X_t}$ is the rms of the corresponding external field. The associated IA stacks are
\begin{equation}
S^{\pm}_{X_tI}(\nu)=
\frac{1}{N^{\pm}_{X_t}(\nu)}
\sum_{p\in{\cal P}^{\pm}_{X_t}(\nu)} I(p),
\label{eq:extrema_stack}
\end{equation}
with $\nu=0,1,2,3$ in all four tracer bins. Extrema are identified independently in each realization.

All higher-order statistics are measured after smoothing the participating fields with the same spherical top-hat window. We adopt $\theta_{\rm TH}=3\,{\rm arcmin}$ as the fiducial scale and repeat the measurements at $12$ and $28\,{\rm arcmin}$ to test the scale dependence. We also verified that the qualitative conclusions are unchanged when the statistics are measured on the unsmoothed maps.

\paragraph{Two-IA statistics.}

Statistics containing two factors of the IA field are intended to probe correlations between the intrinsic shapes of distinct galaxies. A naive product of the pixelized intrinsic-shape field, however, also contains terms in which the same galaxy contributes to both IA factors. These same-object terms measure the one-galaxy ellipticity variance rather than a correlation between different galaxies, and can be substantial for the sparse source samples considered here. We therefore remove them explicitly and construct all two-$I$ observables from distinct-galaxy pairs only.

For each catalogue, we generate eight deterministic random partitions into two disjoint subsets, $A$ and $B$. Defining the full and difference intrinsic-shape fields as
\begin{equation}
I=I_A+I_B,
\qquad
I_\Delta=I_A-I_B,
\end{equation}
where both subset fields retain the normalization of the full source sample, the distinct-object intrinsic product is
\begin{equation}
Q_I(\hat{\boldsymbol n})
=
I^2(\hat{\boldsymbol n})-I_\Delta^2(\hat{\boldsymbol n})
=
4I_A(\hat{\boldsymbol n})I_B(\hat{\boldsymbol n}).
\label{eq:distinct_pair_product}
\end{equation}
Since the two subsets are disjoint, the same galaxy cannot contribute to both intrinsic factors. The resulting estimator therefore excludes self-pairs without requiring an additional subtraction. We average the measurements over the eight partitions.

We use this construction to measure the mixed higher-order statistics
\begin{equation}
GII_t=\left\langle G_t Q_I\right\rangle,
\qquad
dII_t=\left\langle d_t Q_I\right\rangle,
\label{eq:GII_dII}
\end{equation}
in each of the four external tracer bins. The fields are band-limited at $\ell_{\max}=4096$ and smoothed with the same $3\,{\rm arcmin}$ top-hat filter used for the one-$I$ higher-order statistics.

We measure the intrinsic--intrinsic two-point spectra using the harmonic-space analogue of the same distinct-object estimator,
\begin{equation}
C_\ell^{II,{\rm dist}}
=
C_\ell(I,I)
-
\left\langle
C_\ell(I_\Delta,I_\Delta)
\right\rangle_{\rm split},
\label{eq:II_distinct}
\end{equation}
separately for the $EE$ and $BB$ components. These spectra are also restricted to $\ell\leq4096$ but are not smoothed. The $GII$ and $dII$ measurements probe how correlations between intrinsic shapes depend on the surrounding lensing and density fields, while $II$ directly probes pairwise shape coherence and intra-halo geometry. All three two-$I$ statistics are included in the joint calibration together with the one-$I$ observables described above.

\paragraph{Uncertainties and calibration objective.}

For each measurement we define the normalized residual
\begin{equation}
z_i=
\frac{M_i-T_i}{\sigma_{{\rm tot},i}},
\end{equation}
where $T_i$ and $M_i$ denote the FLAMINGO target and model prediction. We estimate the target covariance from 48 spatial delete-one jackknife regions. We also include the model covariance, which typically changes the marginal uncertainties by only about ten per cent. For the independent portable-PKDGRAV realizations, the target and model covariance contributions are added directly in quadrature. For the matched and portable FLAMINGO models, which share the same underlying realization as the target, we instead estimate the covariance of the model--target difference. 

The calibration prioritizes the largest residuals, with $>5\sigma$ outliers weighted most strongly, followed by $>4\sigma$ and $>3\sigma$ excursions. Residual amplitudes are then used to distinguish solutions with similar outlier counts, with equal weighting across statistic families.

We fit all summary statistics jointly with a common set of model parameters. These include the $GI$ and $dI$ spectra, the five one-$I$ mixed moments, the $G$- and $d$-selected extrema stacks, and the distinct-object $II$, $GII$, and $dII$ statistics defined above.

We do not estimate or invert a full covariance matrix across the different statistic families, so the calibration objective is not a likelihood $\chi^2$. The $3\sigma$, $4\sigma$, and $5\sigma$ thresholds are used only as per-bin diagnostics, and correlated bins should not be interpreted as independent significance tests.

\section{Results}
\label{sec:results}

\begin{table*}
  \centering
  \caption{Calibration performance of the IA prescriptions across the full
  fitted data vector, including the one-$I$ and two-$I$ statistics described in
  Sec.~\ref{sec:summary_statistics}. For each sample we report the cumulative
  number of measurements with marginal residuals exceeding $3\sigma$, $4\sigma$,
  and $5\sigma$, respectively. The fitted vectors contain 219, 227, and 226
  valid measurements for RED, BLUE, and Euclid-like (the number changes depending on the valid entries for the peak and minima statistics). These counts are
  descriptive calibration diagnostics; because the measurements are correlated,
  they should not be interpreted as numbers of independent significances.}
  \label{tab:model_performance}
  \begin{tabular}{lccc}
  \toprule
  Model
  & RED
  & BLUE
  & Euclid-like \\
  \midrule
  NLA
  & 104/83/67
  & 107/79/65
  & 133/98/77 \\

  $\delta$NLA
  & 92/71/59
  & 91/75/55
  & 109/80/67 \\

  Halo, matched
  & 0/0/0
  & 0/0/0
  & 0/0/0 \\

  Halo, portable FLAMINGO
  & 0/0/0
  & 0/0/0
  & 0/0/0 \\

  Halo, portable PKDGRAV
  & 14/0/0
  & 3/0/0
  & 2/0/0 \\
  \bottomrule
  \end{tabular}
  \end{table*}

\begin{figure*}[t]
    \centering
    \includegraphics[width=\textwidth, height=0.82\textheight,keepaspectratio]
    {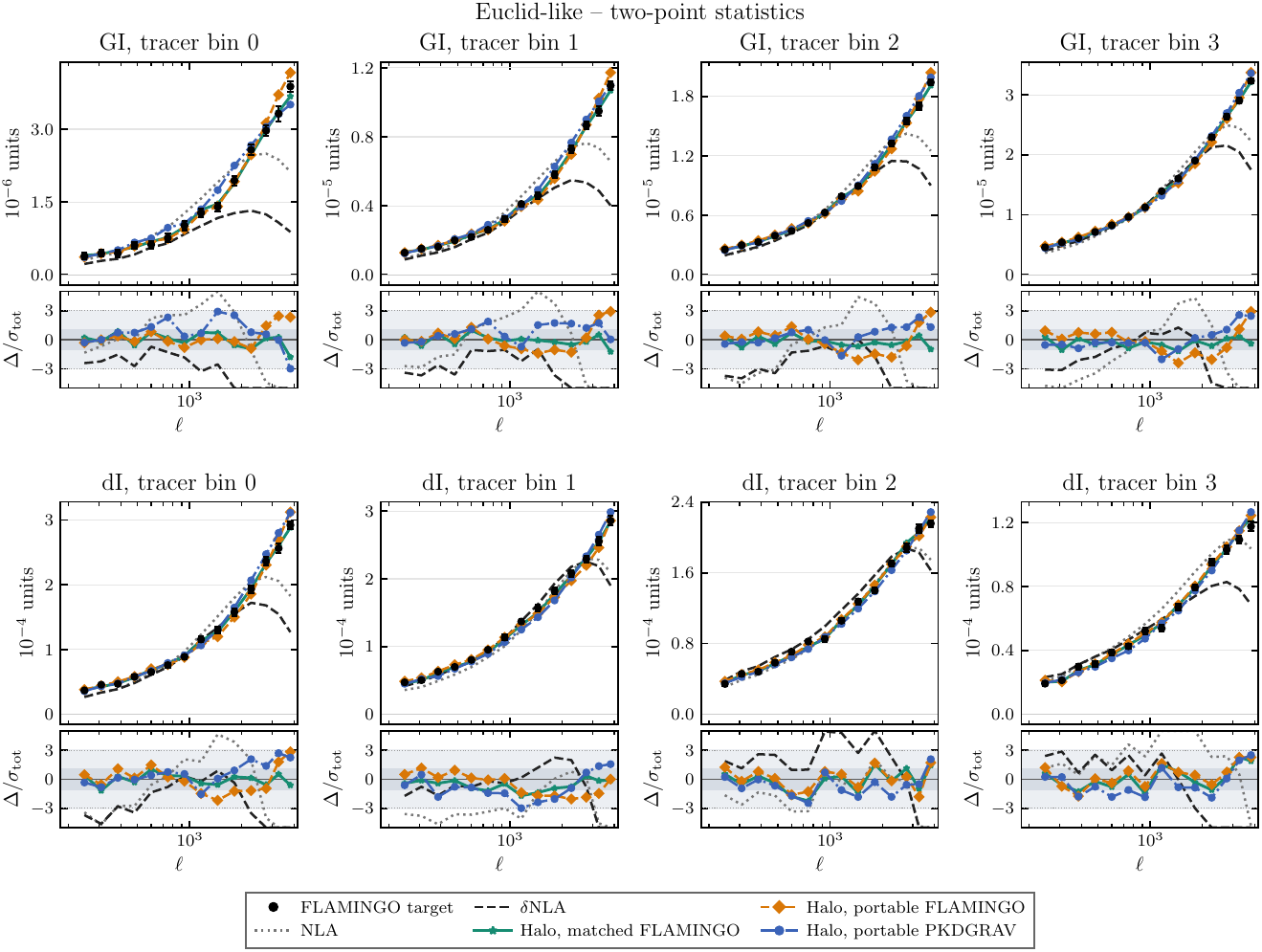}
    \caption{
    Two-point IA statistics for the Euclid-like sample. The upper panels show the $GI$ (top row) and $dI$ (bottom row) band powers in the four tracer bins. Black points show the FLAMINGO target, with error bars corresponding to the target uncertainty only; grey and black dashed curves show the NLA and $\delta$NLA prescriptions, and the coloured curves show the matched and portable halo models. The lower sub-panels show residuals relative to FLAMINGO in units of $\sigma_{\rm tot}$, which also includes the model uncertainty and therefore depends on the model construction; this typically changes the marginal uncertainty by about ten per cent. All model parameters are fitted jointly to the full set of summary statistics rather than independently to each panel. The halo model provides a substantially better simultaneous description of the nonlinear scale dependence than the tidal-field prescriptions.
    }
    \label{fig:euclid_twopoint}
\end{figure*}

\begin{figure*}[t]
    \centering
    \includegraphics[width=\textwidth, height=0.82\textheight,keepaspectratio]
    {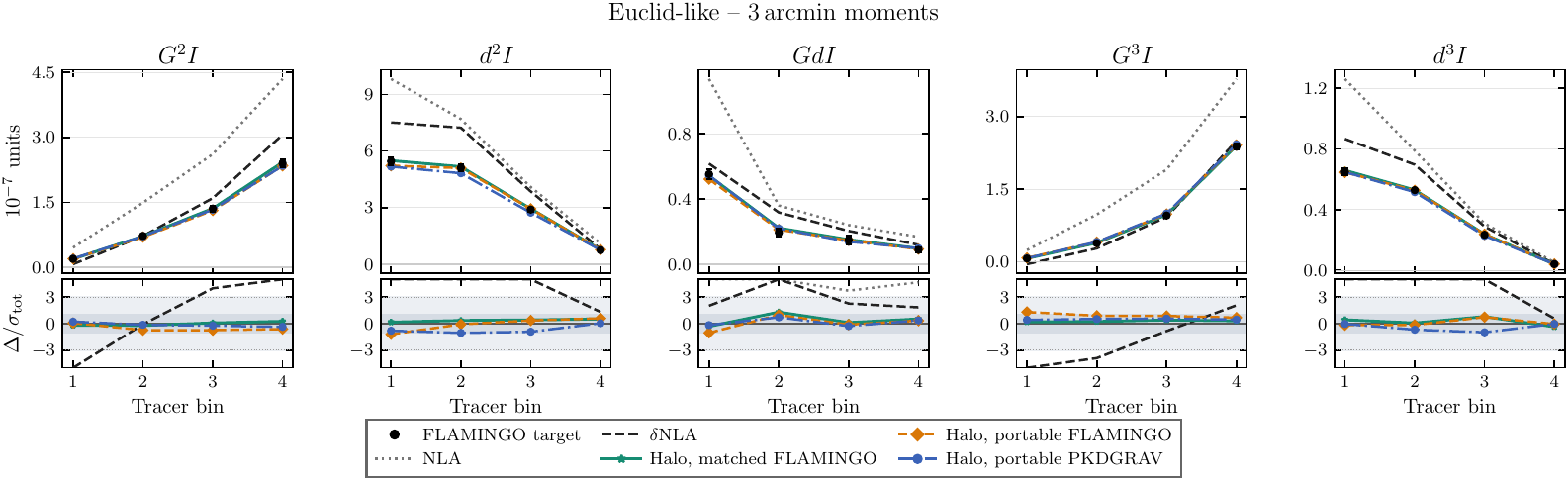}
    \caption{
    One-IA mixed moments for the Euclid-like sample, measured after smoothing all fields with a $3\,{\rm arcmin}$ spherical top-hat. From left to right we show $\langle G^2 I\rangle$, $\langle d^2 I\rangle$, $\langle GdI\rangle$, $\langle G^3I\rangle$, and $\langle d^3I\rangle$ in the four tracer bins. Symbols, line styles, and the residual convention follow Fig.~\ref{fig:euclid_twopoint}. The halo model reproduces the mixed moments across the different tracer combinations, whereas NLA and $\delta$NLA show coherent departures in several statistic families.
    }
    \label{fig:euclid_moments}
\end{figure*}

\begin{figure*}[t]
    \centering
    \includegraphics[width=\textwidth, height=0.82\textheight,keepaspectratio]
    {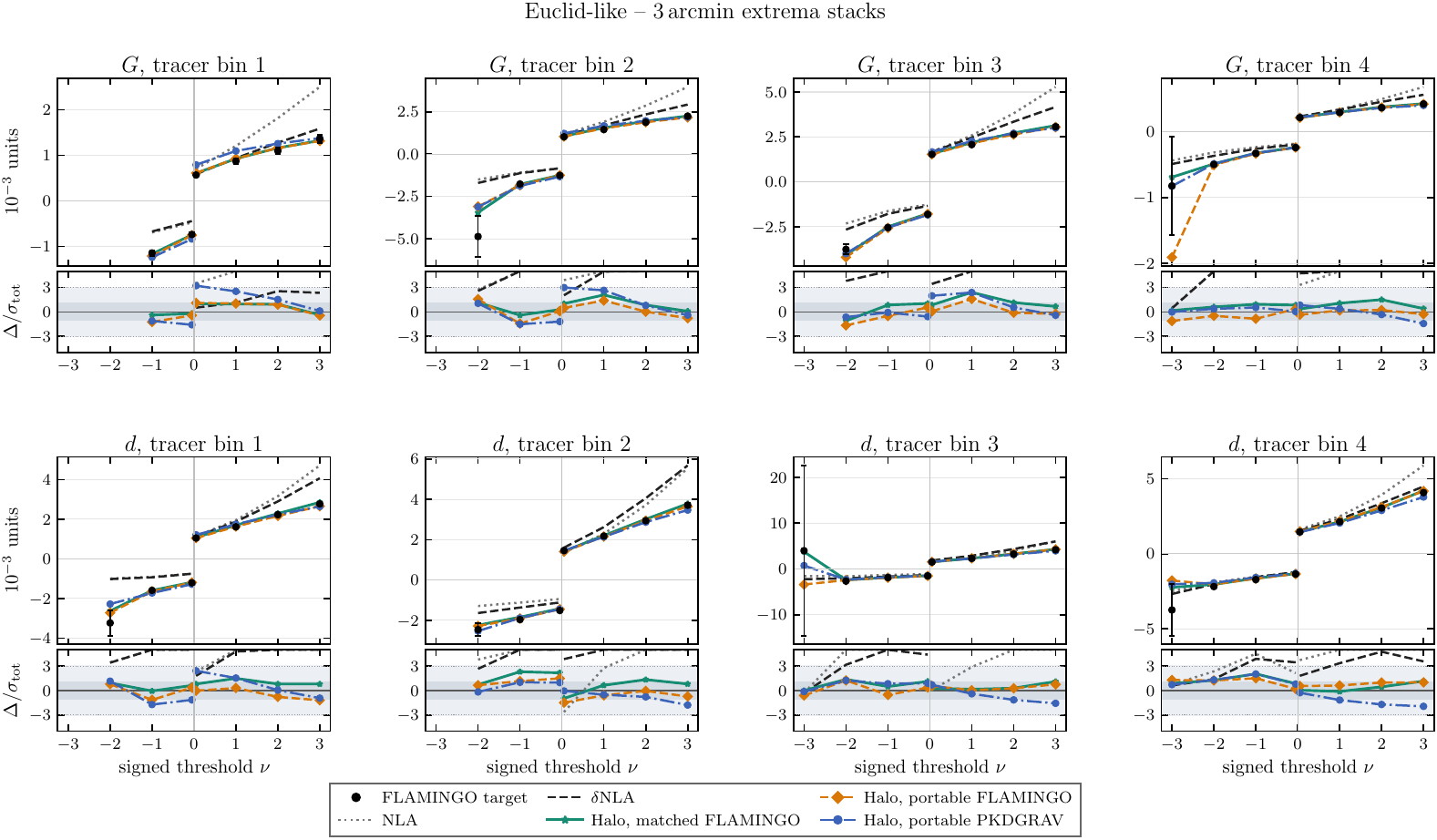}
    \caption{
    IA stacks around extrema of the external fields for the Euclid-like sample, using $3\,{\rm arcmin}$ top-hat smoothing. The top row shows extrema selected in the lensing field $G$, while the bottom row shows extrema selected in the projected density field $d$, for the four tracer bins. Positive threshold values correspond to maxima and negative values to minima. Symbols and residuals follow the convention of Fig.~\ref{fig:euclid_twopoint}. Missing points at large negative threshold indicate cases in which no sufficiently deep minima are present. The halo model provides a substantially better joint description of the extrema-conditioned IA signal than the tidal-field prescriptions.
    }
    \label{fig:euclid_extrema}
\end{figure*}

\begin{figure*}[t]
    \centering
    \includegraphics[width=\textwidth, height=0.82\textheight,keepaspectratio]
    {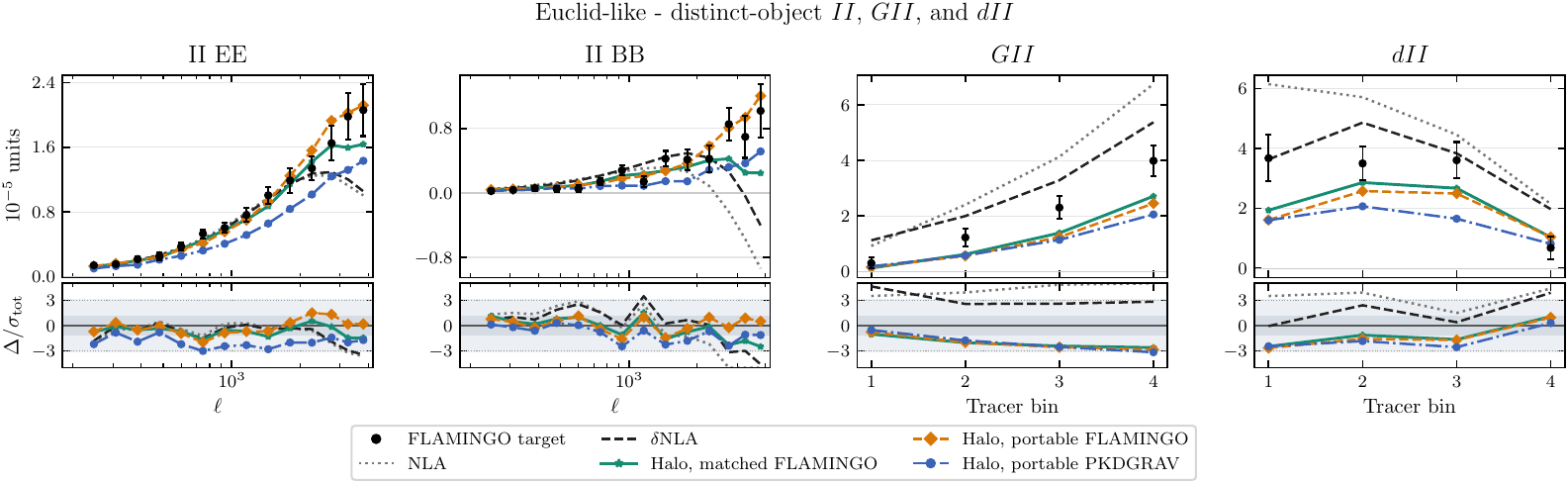}
    \caption{
    Distinct-object two-IA statistics for the Euclid-like sample. The first two panels show the $II$ $EE$ and $BB$ power spectra, while the remaining panels show the $GII$ and $dII$ statistics in the four external tracer bins. Same-galaxy self-pairs are removed using the split-catalogue estimator described in Sec.~\ref{sec:summary_statistics}. Symbols, line styles, and residuals follow the convention of Fig.~\ref{fig:euclid_twopoint}. These observables are fitted jointly with the one-IA statistics and provide additional sensitivity to pairwise intrinsic-shape correlations and intra-halo central--satellite structure. 
    }
    \label{fig:euclid_II}
\end{figure*}

In this section we evaluate the performance of the IA prescriptions against the FLAMINGO target across the full set of summary statistics. We first isolate the intrinsic-alignment response by evaluating the three IA prescriptions at the positions of the original FLAMINGO galaxies. We then remove the galaxy catalogue and regenerate the source population from the halo field, first using the FLAMINGO halo catalogue and subsequently an independent PKDGRAV realization. In all cases, the model parameters are calibrated jointly to the full set of one- and two-IA statistics described in Sec.~\ref{sec:summary_statistics}. We show the Euclid-like sample in the main figures and use the RED and BLUE samples as complementary tests of substantially different galaxy populations; selected additional figures are collected in Appendix~\ref{app:sample_results}, while their quantitative performance is summarized below.

\subsection{Matched FLAMINGO: testing the IA response}
\label{sec:results_matched}

Figure~\ref{fig:euclid_twopoint} shows the matched $GI$ and $dI$ spectra for the Euclid-like sample, Figs.~\ref{fig:euclid_moments} and~\ref{fig:euclid_extrema} show the corresponding one-IA higher-order statistics, and Fig.~\ref{fig:euclid_II} shows the distinct-object $II$, $GII$, and $dII$ measurements. All of these observables are fitted simultaneously with a single set of model parameters. We summarize the overall quality of the fit through the number of measurements with residuals exceeding $3\sigma$, $4\sigma$, and $5\sigma$, reported in Table~\ref{tab:model_performance}.

The figures and Table~\ref{tab:model_performance} show that NLA captures the broad sign and redshift evolution of the alignment signal and describes substantial parts of the two-point measurements on large and intermediate scales, but significant discrepancies remain once nonlinear scales and higher-order observables are fitted simultaneously. Adding the density-weighted contribution in $\delta$NLA reduces the overall number of outliers, but still does not provide a satisfactory joint description. In some individual panels it can even perform worse than NLA, despite its additional freedom, because the parameters are optimized across all statistics and scales rather than for each observable separately. The mismatch is therefore not confined to a particular multipole or statistic, but reflects the difficulty of reproducing the full scale and statistic dependence of the simulated IA field with these tidal-field prescriptions. Since the matched models are evaluated at the positions of the actual FLAMINGO galaxies, this discrepancy cannot be attributed to the source-population model and instead points to limitations of the IA response itself. 

By contrast, the matched halo model provides a good simultaneous description across the full set of observables, which probe complementary aspects of the IA field. The $GI$ and $dI$ spectra measure the cross-correlation of intrinsic shapes with the lensing and projected-density fields, which weight the matter distribution with different radial kernels. The mixed higher-order moments probe nonlinear couplings between intrinsic shape, density, and lensing, while the extrema stacks test the IA response in strongly over- and under-dense projected environments. Across these one-IA observables, the halo model closely follows the FLAMINGO measurements. The distinct-object $II$, $GII$, and $dII$ statistics extend this comparison to correlations between the intrinsic shapes of different galaxies. Although noisier, they are more directly sensitive to pairwise intra-halo structure, including central--satellite and satellite--satellite contributions and their relative geometry, which is not directly constrained by the one-IA statistics. As shown in Fig.~\ref{fig:euclid_II}, the matched halo model reproduces these observables as well. The same prescription therefore captures both the response of intrinsic shapes to the surrounding matter field and the coherence of intrinsic shapes between galaxies.

This success is especially informative in the matched construction because the galaxy population, central--satellite classification, and host assignments are fixed. The improvement therefore arises from the IA response itself, namely the separate central and satellite contributions tied to halo geometry and environment. This conclusion holds across all three source samples. As summarized in Table~\ref{tab:model_performance}, the matched halo model leaves no measurements above $3\sigma$ for RED, BLUE, or Euclid-like, compared with $\mathcal{O}(100)$ large residuals for NLA and $\delta$NLA. These counts are correlated diagnostics rather than numbers of independent statistical excursions, but they provide a compact summary of the much better simultaneous description achieved by the halo model. Since the tidal-field prescriptions already fail in this matched setting, where the source population is exact, we do not pursue portable versions of NLA or $\delta$NLA and restrict the source-population and transfer tests below to the halo model.

\subsection{Regenerating the source population}
\label{sec:results_portable_flamingo}

\begin{figure}
    \centering
    \includegraphics[width=0.45\textwidth]
    {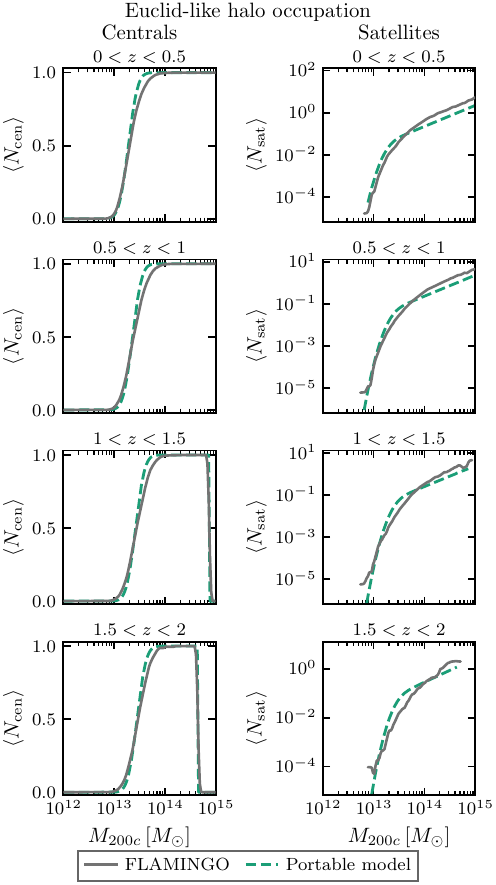}
    \caption{
    Halo occupation of the Euclid-like source sample in four redshift
    intervals. The upper panels show the mean central occupation and the
    lower panels the mean satellite occupation as a function of
    $M_{200c}$. Black curves show the occupation measured from the
    shape-selected FLAMINGO sample, while dashed curves show the minimal
    abundance-conditioned portable population model. The model is not
    intended as a precision HOD fit; its purpose is to reproduce the broad
    halo-mass and redshift dependence of the selected population while
    keeping the number of additional nuisance parameters small.
    }
    \label{fig:euclid_hod}
\end{figure}

We next remove the direct use of the FLAMINGO galaxy catalogue and generate the source population from the halo catalogue using the prescription of Sec.~\ref{sec:halo_ia}. The portable model introduces the two population parameters $f_c$ and $f_{\rm sat}$, while the remaining occupation quantities are fixed by the abundance constraints. The aim is not to reproduce the detailed halo occupation, but to capture the broad host-mass and redshift dependence of the galaxies that dominate the IA signal.

Figure~\ref{fig:euclid_hod} compares the resulting occupation model with the Euclid-like FLAMINGO population. The simple prescription reproduces the broad central and satellite trends, and the corresponding two-point, higher-order, and two-IA statistics remain close to the matched halo predictions in Figs.~\ref{fig:euclid_twopoint}--\ref{fig:euclid_II}. As shown in Table~\ref{tab:model_performance}, the portable FLAMINGO model leaves no measurements above $3\sigma$ for any of the three source samples. The quality of the fit therefore remains essentially unchanged after the FLAMINGO galaxy positions and host assignments are removed.

The RED sample provides the more stringent test of the population model. Its central occupation is strongly non-monotonic and is poorly represented by the adopted HOD error-function form, particularly at low redshift, yet the IA statistics remain well reproduced. This indicates that a precise reconstruction of the full HOD is not required for the observables considered here; capturing the broad halo populations that dominate the IA-weighted field appears to be sufficient. The RED occupation and the RED and BLUE two-point measurements are shown in Appendix~\ref{app:sample_results}.

\subsection{Transfer to an independent PKDGRAV realization}
\label{sec:results_pkdgrav}

We next apply the same architecture to the independent PKDGRAV realization. This changes the nonlinear density realization, the halo finder, the mass convention, and the halo inertia tensors, while removing any dependence on the identities or positions of the FLAMINGO galaxies. The population and IA-response parameters are recalibrated, but the model architecture is otherwise unchanged.

The transfer remains successful across the full fitted data vector. As shown in Table~\ref{tab:model_performance}, the portable PKDGRAV model leaves $14/0/0$, $3/0/0$, and $2/0/0$ measurements above $3\sigma/4\sigma/5\sigma$ for the RED, BLUE, and Euclid-like samples, respectively. These correspond to only a small fraction of the more than 200 measurements in each sample, and no residual exceeds $4\sigma$. The residual counts are larger than for the matched and portable FLAMINGO models, for which no $>3\sigma$ residuals remain after calibration, but this comparison should not be interpreted probabilistically: the residuals are correlated and the calibration objective explicitly prioritizes the suppression of large outliers. More importantly, the PKDGRAV result tests the model on an independent density realization and halo catalogue. The remaining discrepancies are therefore modest given the change in simulation and halo construction, and are distributed across the fitted data vector rather than indicating a systematic failure of a particular class of observables. This indicates that the halo prescription is not strongly tied to the specific halo definition, mass convention, or density realization used in FLAMINGO.

\begin{table*}
\centering
\caption{Parameter-restriction tests for the halo IA model. For each restricted model we re-fit all remaining free parameters and report the cumulative number of measurements with residuals exceeding $3\sigma$, $4\sigma$, and $5\sigma$. Results are shown for the matched FLAMINGO construction and for portable FLAMINGO, which additionally regenerates the source population from the halo catalogue. Parameters listed in parentheses are fixed to zero.}
\label{tab:halo_parameter_tests}
\begin{tabular}{lcc cc cc}
\toprule
& \multicolumn{2}{c}{RED}
& \multicolumn{2}{c}{BLUE}
& \multicolumn{2}{c}{Euclid-like} \\
\cmidrule(lr){2-3}
\cmidrule(lr){4-5}
\cmidrule(lr){6-7}
Restriction
& Matched & Portable
& Matched & Portable
& Matched & Portable \\
\midrule
Full model
& 0/0/0 & 0/0/0
& 0/0/0 & 0/0/0
& 0/0/0 & 0/0/0 \\

No central IA ($A_{c,0}=0$)
& 133/119/95 & 139/119/99
& 165/149/133 & 184/167/150
& 170/147/133 & 183/167/151 \\

No satellite IA ($A_{s,0}=0$)
& 74/42/30 & 24/8/1
& 21/10/1 & 12/5/0
& 41/20/12 & 45/20/10 \\

No density dependence ($b_{{\rm TA},c}=b_{{\rm TA},s}=0$)
& 8/0/0 & 0/0/0
& 0/0/0 & 10/1/0
& 0/0/0 & 0/0/0 \\

No central density dependence ($b_{{\rm TA},c}=0$)
& 4/0/0 & 0/0/0
& 0/0/0 & 4/0/0
& 0/0/0 & 0/0/0 \\

No satellite density dependence ($b_{{\rm TA},s}=0$)
& 1/0/0 & 0/0/0
& 0/0/0 & 2/0/0
& 0/0/0 & 0/0/0 \\

No central redshift evolution ($\eta_c=0$)
& 6/0/0 & 26/5/1
& 0/0/0 & 1/0/0
& 0/0/0 & 0/0/0 \\

No satellite redshift evolution ($\eta_s=0$)
& 1/0/0 & 0/0/0
& 0/0/0 & 1/0/0
& 5/0/0 & 1/0/0 \\

No redshift evolution ($\eta_c=\eta_s=0$)
& 8/1/0 & 32/4/1
& 0/0/0 & 2/0/0
& 2/0/0 & 0/0/0 \\

Amplitudes only ($\eta_c=\eta_s=b_{{\rm TA},c}=b_{{\rm TA},s}=0$)
& 9/1/0 & 41/18/4
& 0/0/0 & 6/0/0
& 2/0/0 & 3/0/0 \\
\bottomrule
\end{tabular}
\end{table*}
\subsection{Which halo-model ingredients matter?}
\label{sec:results_ablation}

We next ask whether the halo model can be simplified further. Starting from the full six-parameter IA response, we fix selected terms to zero and re-fit the remaining parameters. We do this for both the matched and portable FLAMINGO constructions, so that we can distinguish response complexity that is already required at the true galaxy positions from extra flexibility introduced when the source population is regenerated. The resulting residual counts are reported in Table~\ref{tab:halo_parameter_tests}.

The first clear result is that the separate central and satellite amplitudes are both required. Setting either $A_{c,0}$ or $A_{s,0}$ to zero produces many large residuals already in the matched case. By contrast, most of the redshift- and density-dependent terms can be removed with only a small degradation.

For BLUE and Euclid-like, this reduction is particularly strong. In the matched construction, the two amplitudes alone leave $0/0/0$ and $2/0/0$ residuals above $3\sigma/4\sigma/5\sigma$, respectively. After regenerating the galaxy population, the corresponding four-parameter portable model, $\{f_c,f_{\rm sat},A_{c,0},A_{s,0}\}$, still performs very well, with $6/0/0$ residuals for BLUE and $3/0/0$ for Euclid-like. For these samples there is therefore little evidence that additional redshift or density dependence is required.

RED is the main exception. Even in the matched case, amplitudes alone leave $9/1/0$ residuals, and the degradation becomes much stronger in the portable construction, where the same restriction gives $41/18/4$. The central redshift-evolution parameter is especially important: fixing $\eta_c=0$ gives $6/0/0$ in the matched case but $26/5/1$ after regenerating the population. This is consistent with the larger mismatch between the target and portable RED central populations, which grows with redshift. In this case, $\eta_c$ appears to absorb both genuine evolution of the RED IA response and imperfections in the portable HOD.

Overall, the results suggest that the core model is quite compact: two IA amplitudes are sufficient in the matched case for BLUE and nearly sufficient for Euclid-like, while the portable model generally requires only those two amplitudes plus the two population parameters. Additional response terms become useful mainly when the source-population model is imperfect, with RED providing the clearest example.

\section{Discussion}
\label{sec:discussion}

The results above show that a relatively simple halo-based construction can reproduce a broad range of IA observables while remaining applicable after the original hydrodynamical galaxy catalogue is removed. In particular, the matched tests provide the cleanest comparison between the different IA prescriptions because the galaxy positions, central--satellite classifications, and host assignments are fixed to their FLAMINGO values. In this setting, NLA and $\delta$NLA reproduce parts of the large- and intermediate-scale two-point signal but fail to describe simultaneously the nonlinear $GI$ and $dI$ spectra, the higher-order statistics, and the two-IA observables. The failure is therefore already present at the level of the IA response and cannot be attributed to an approximate model for the source population. The halo prescription, by contrast, provides a common description across all of these observables and across the RED, BLUE, and Euclid-like samples.

We have not tested a full TATT implementation with independent tidal-torquing terms. The $\delta$NLA model considered here is instead a phenomenological field-level extension of NLA, constructed from the nonlinear projected density and tidal fields. More complete tidal prescriptions may improve the description over some of the scales considered here. Our comparison should therefore not be interpreted as a general test of all tidal-alignment models, but as evidence that the simple field-level prescriptions considered here do not capture the full nonlinear IA structure seen in FLAMINGO.

The fact that the halo model simultaneously reproduces both one-IA observables and statistics involving correlations between distinct galaxy shapes suggests that its central--satellite decomposition captures relevant intra-halo structure beyond the mean response to the surrounding matter field. For centrals, the projected dark-matter halo tensor acts as a coherent template rather than a deterministic prediction of the stellar shape. Hydrodynamical simulations show substantial scatter between galaxy and halo orientations and ellipticities \citep{Shao2016,Chisari2017}; in our model this is absorbed at leading order by the central response amplitude while retaining the preferred halo orientation and its variation between objects. For satellites, the model combines radial intrinsic alignments with a spatial distribution that follows the projected halo geometry. These ingredients provide a simple way of explicitly modelling the intra-halo central--satellite structure that is not present in the tidal-field prescriptions considered here. The broad architecture is closely related to the empirical HOD+IA framework of \citet{VanAlfen2024} and its recent emulator and differentiable extensions \citep{Pandya2025IAEmu,Pandya2026diffHODIA}. Those studies primarily calibrate and validate the model using one- and two-point galaxy and IA statistics, whereas here we test the same class of halo-based description jointly across two-point, higher-order, and two-IA observables measured directly from projected fields.

An important result is that the model does not require a detailed source-population prescription. After removing the FLAMINGO galaxy catalogue and regenerating the population directly from the halo field, the portable FLAMINGO model retains the same level of agreement with the target for all three samples. This remains true even for RED, whose strongly non-monotonic central occupation is not reproduced accurately by the simple HOD adopted here. The relevant requirement therefore appears to be to capture the broad halo populations contributing to the IA field rather than every detail of the galaxy occupation.

The parameter-restriction tests show that the model can also be reduced considerably. Separate central and satellite amplitudes are the main required IA ingredients, while the redshift- and density-response terms are much less important for BLUE and Euclid-like. In the portable case, this leaves a compact four-parameter model, $\{A_{c,0},A_{s,0},f_c,f_{\rm sat}\}$, which has the same number of free parameters as $\delta$NLA but provides a much better simultaneous description of the fitted statistics. RED requires additional central redshift dependence, which appears to capture both genuine evolution of the RED IA response and imperfections in the regenerated source population.

The transfer to PKDGRAV provides a more stringent test because the density realization, halo finder, halo mass definition, and inertia tensors all change. Despite these differences, the same model architecture continues to reproduce the full set of fitted statistics. The PKDGRAV comparison is therefore the more relevant test of whether the model generalizes beyond the realization on which it was developed, and its performance suggests that the prescription is not strongly tied to a particular halo catalogue or density field.

The next step is to test the same construction across additional hydrodynamical simulations and galaxy selections, particularly at lower halo masses and for more realistic Stage-IV source samples. The Euclid-like sample used here is defined only through its redshift distribution and does not include a detailed survey selection. A survey-level forward model will also require a stochastic intrinsic-shape component together with observational effects such as measurement noise, blending, and selection.

Finally, the discrepancies found for NLA and $\delta$NLA do not imply that these models are necessarily inadequate for current cosmic-shear analyses. Our tests deliberately use low-noise simulated fields and include nonlinear and non-Gaussian observables designed to expose differences between IA prescriptions. Whether these differences are detectable, or lead to significant cosmological biases, for a realistic survey selection and covariance is a separate question. The purpose of the present work is instead to establish a practical field-level IA model for forward simulations in which the same intrinsic-shape realization can be propagated consistently through two-point and non-Gaussian analyses.

\section{Summary and outlook}
\label{sec:summary}

We have developed and tested a halo-based prescription for generating intrinsic galaxy shapes in field-level weak-lensing simulations. The model separates central and satellite galaxies, uses projected dark-matter halo shapes to define the central IA template, assigns radial alignments to satellites, and generates the source population from a low-dimensional halo occupation model. We calibrate and test the model against intrinsic galaxy shapes measured in FLAMINGO using a common set of two-point and higher-order statistics involving the IA, shear, and density fields, and compare its performance with field-level NLA and density-weighted NLA prescriptions. We consider three galaxy samples---RED, BLUE, and Euclid-like---to test the model across populations with different redshift distributions and IA properties.

Our main results are:

\begin{itemize}

\item When evaluated at the positions of the original FLAMINGO galaxies, so that the galaxy--matter connection is fixed exactly, the halo model provides a simultaneous description of the full set of IA statistics for all three samples. In contrast, NLA and $\delta$NLA leave substantial discrepancies, particularly on nonlinear scales and in the higher-order observables.

\item The agreement of the halo model is preserved when the FLAMINGO galaxy catalogue is removed and the source population is regenerated directly from the halo catalogue. The portable halo model continues to reproduce the measurements for all three samples using a simple low-dimensional HOD. The RED sample provides the most stringent case: its halo occupation is not reproduced in detail, yet this does not prevent an accurate description of the IA statistics.

\item We also tested reduced versions of the halo model to determine how much freedom is actually required. Our results suggest that a four-parameter portable model---two IA amplitudes and two population parameters---may already be sufficient in many cases. Additional freedom becomes useful when the source population is more difficult to represent with the simplified HOD, and can enter either through a more flexible population model or through extra IA-response parameters that effectively compensate for residual mismatches.

\item The same halo-model architecture also transfers successfully to an independent $N$-body realization evolved with a different gravity-only code (\textsc{PKDGRAV3}), despite differences in the halo definition, mass convention, and inertia tensors, while retaining a good description of the fitted observables.

\end{itemize}

These results are encouraging for simulation-based inference, where IA must be propagated at field level across a broad range of nonlinear observables. They suggest that moving beyond simple NLA, $\delta$NLA, or similarly low-order tidal prescriptions may be necessary, but that this need not come at the cost of a large nuisance-parameter space: in the simplest cases, the portable halo model is already effective with only four free parameters. Before applying this framework to real survey data, however, it will be important to extend the validation to lower halo masses, test the model against additional hydrodynamical simulations, and repeat these tests with realistic survey selections and observational effects. A survey-level implementation will also require the coherent IA field developed here to be combined with a stochastic intrinsic-shape component and incorporated into full forward simulations.

\begin{acknowledgments}
This research used resources of the National Energy Research Scientific Computing Center, a DOE Office of Science User Facility supported by the Office of Science of the U.S. Department of Energy under Contract No. DE-AC02-05CH11231 using NERSC award HEP-ERCAP0037257.
Marco Gatti acknowledges support from the Ramón y Cajal program under grant RYC2024-049031-I, funded by MICIU/AEI/10.13039/501100011033 and by the European Social Fund Plus (FSE+). Marco Gatti thanks Simone Vinciguerra, Supranta Boruah, Christos Georgiou and Jonathan Blazek for useful discussions and comments on the manuscript.
\end{acknowledgments}

\bibliographystyle{apsrev4-2}
\bibliography{bibliography}

\appendix

\section{Results for the colour-selected samples}
\label{app:sample_results}

The main text focuses on the Euclid-like sample. Here we show additional results for the RED and BLUE colour-selected samples, which probe substantially different galaxy populations and redshift distributions.

\begin{figure*}
    \centering
    \includegraphics[width=\textwidth]
    {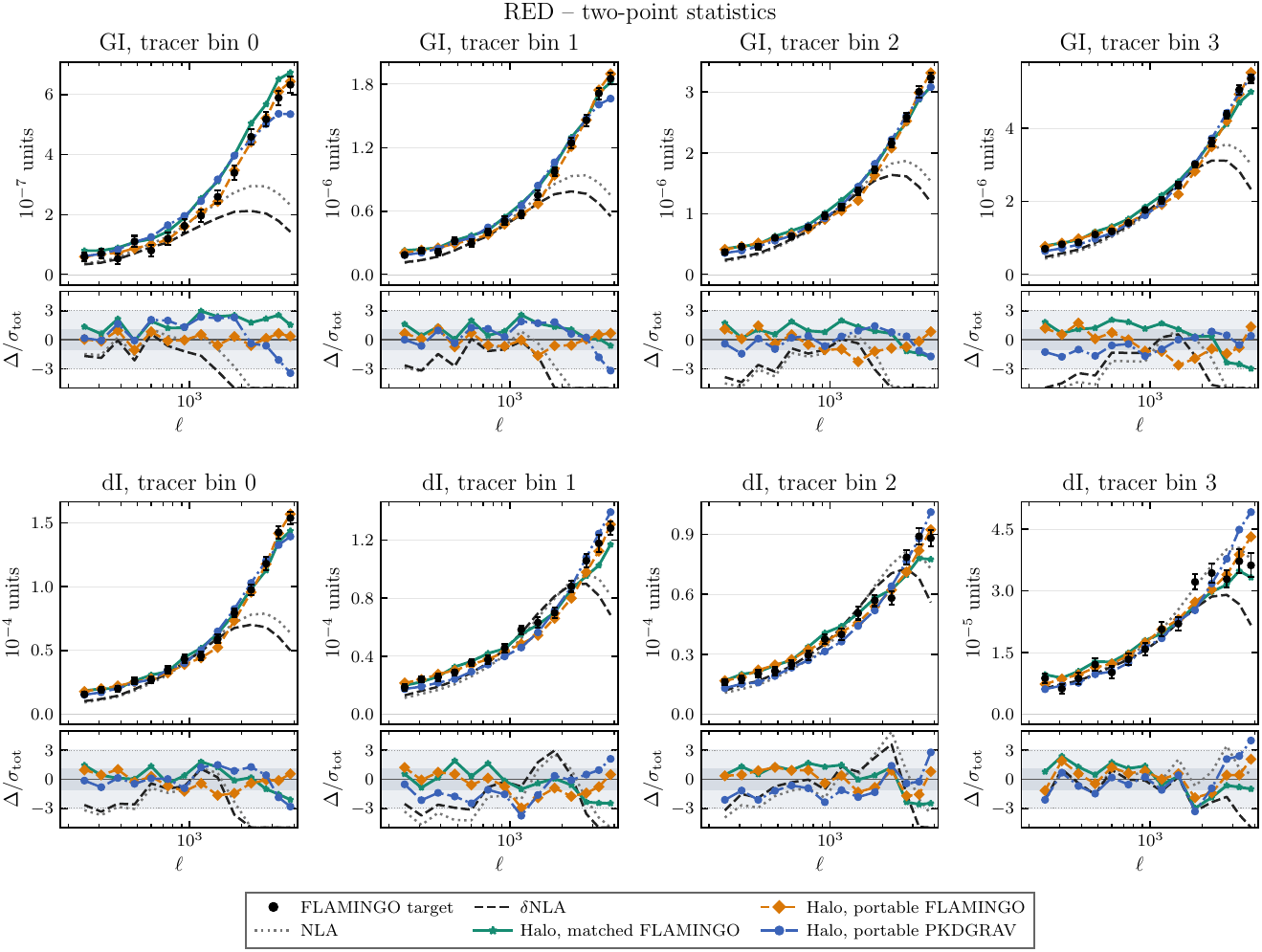}
    \caption{
    Two-point $GI$ and $dI$ statistics for the RED sample. The layout, symbols, and residual convention follow Fig.~\ref{fig:euclid_twopoint}. The matched and portable halo models provide a substantially better simultaneous description of the nonlinear scale dependence than NLA and $\delta$NLA.
    }
    \label{fig:red_twopoint}
\end{figure*}

\begin{figure*}
    \centering
    \includegraphics[width=\textwidth]
    {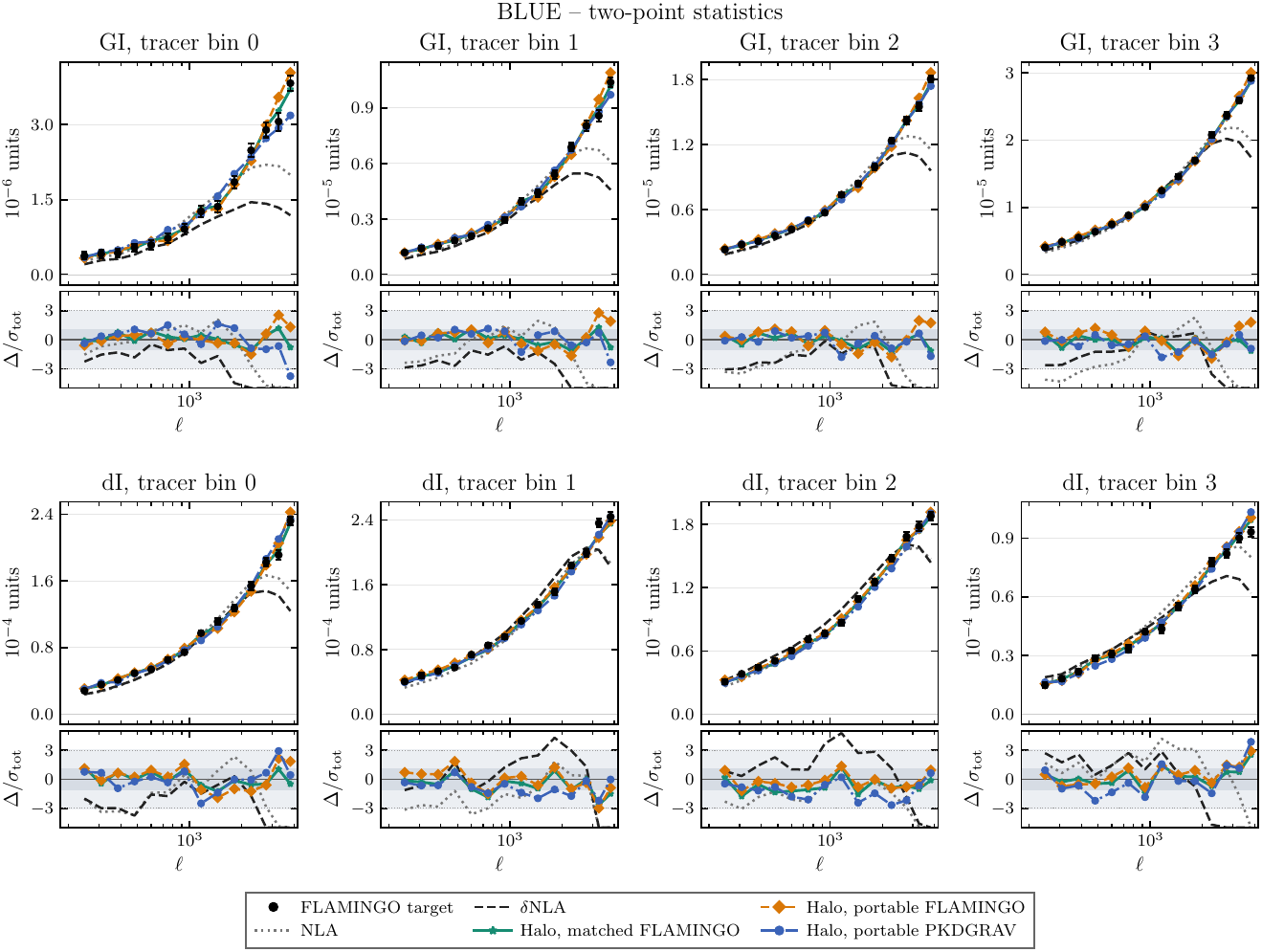}
    \caption{
    Same as Fig.~\ref{fig:red_twopoint}, for the BLUE sample. The matched and portable halo models remain close to the FLAMINGO target across the four tracer bins.
    }
    \label{fig:blue_twopoint}
\end{figure*}

Figures~\ref{fig:red_twopoint} and~\ref{fig:blue_twopoint} show the corresponding $GI$ and $dI$ measurements. The conclusions are unchanged: NLA and $\delta$NLA do not provide a satisfactory simultaneous description of the nonlinear spectra, while the matched and portable halo models remain close to the FLAMINGO target across the four tracer bins.

\begin{figure}
    \centering
    \includegraphics[width=0.45\textwidth]
    {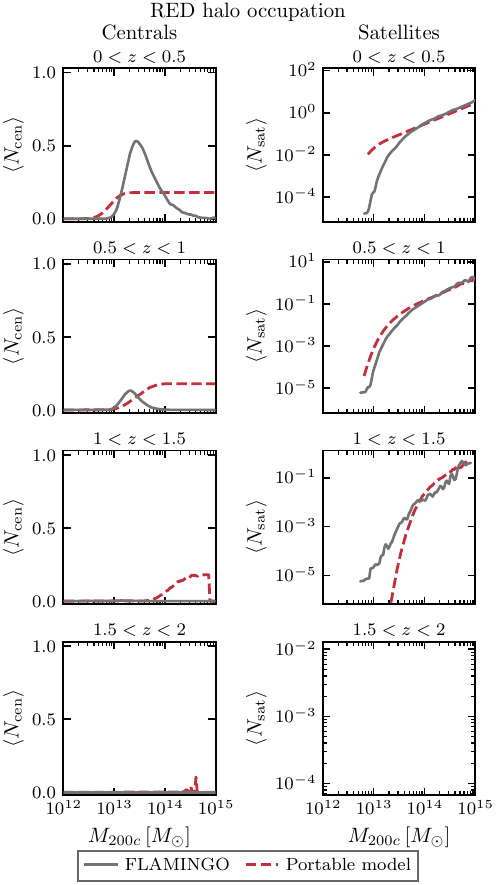}
    \caption{
    Halo occupation of the RED sample. Black curves show the occupation measured from FLAMINGO and dashed curves the portable population model. The central occupation is strongly non-monotonic and is not captured in detail by the simple parameterization, particularly at low redshift, making RED a useful stress test of the source-population model.
    }
    \label{fig:red_hod}
\end{figure}

The RED sample provides the more stringent test of the portable source-population model. Its central occupation is strongly non-monotonic (Fig. \ref{fig:red_hod}) and is not reproduced in detail by the simple HOD adopted in Sec.~\ref{sec:halo_ia}. We deliberately retain the same low-dimensional population model rather than introducing sample-specific freedom. Despite this mismatch, the portable halo model continues to reproduce the IA statistics, showing that an exact reconstruction of the detailed occupation is not required for the observables considered here.

\end{document}